\documentclass[pra,twocolumn,amsmath,showpacs]{revtex4}
\usepackage{graphics}
\begin{document}
\title{A quantum beam splitter for atoms.}

\author{Uffe V. Poulsen} 
\email{uvp@ifa.au.dk}
\author{Klaus M{\o}lmer}
\affiliation{ Institute of Physics and Astronomy, University of
  Aarhus, DK-8000 \AA rhus C}

\begin{abstract}
An interferometric method is proposed to controllably split an atomic 
condensate in two spatial components with strongly reduced population 
fluctuations.  All steps in our proposal are in current use in cold 
atom laboratories, and we show with a theoretical calculation that
our proposal is very robust against imperfections of the interferometer.
\end{abstract}


\maketitle

\section{Introduction}
Cold atoms are routinely applied in spectroscopy and in atomic clocks,
and matter wave interferometers have been used to probe with
unprecedented precision the effects caused by acceleration, gravity or
interaction with atoms or electromagnetic
fields~\cite{berman97:_atom_inter}.  The experimental resolution in
these experiments is linked to the precision with which the relevant
signal can be measured and to the intrinsic uncertainty of the
quantity measured. For example, in Ramsey interferometry applied in
atomic clocks, one tunes a radiation frequency to obtain a 50 \%
population of both states of the clock transition, but if the atoms
are uncorrelated, in addition to the measurement errors one will
inevitably observe differences scaling as $\sqrt{N}$ between the
populations of the two states, where $N$ is the total number of atoms
(binomial distribution). So-called spin squeezed states have been
shown to exist, for which it is possible to significantly reduce this
uncertainty without going to exceedingly large numbers of atoms
\cite{wineland94:_squeez,agarwal90:_cooper}. A number of proposals
has been made to produce such spin squeezed states by interaction with
squeezed light
\cite{kuzmich97:_spin_squeez_ensem_atoms_illum_squeez_light,hald99:_spin_squeez_atoms}, by quantum non-demolition measurement of
the atomic populations
\cite{kuzmich00:_gener_spin_squeez_contin_quant_nondem_measur,moelmer99:_twin},
and by light-induced
\cite{bouchoule01:_spin_squeez_atoms_rydber_block} or collisionally
induced
\cite{soerensen01:_many_bose,dunningham01:_relat_bose_einst,raghavan01:_gener_dicke_bose_einst}
interaction.

In this paper we propose a method to realize a beam splitter for a
one-component condensate, causing a splitting of the condensate in two
spatially separated components with a better matching of occupancies
than in the binomial distribution, resulting from a normal splitting
of the particles. Such splitting has indeed been proposed to occur if
one adiabatically raises a potential barrier inside a single
condensate, and experiments have confirmed the effects of collisional
interactions on the population statistics of a condensate split by a
periodic potential \cite{orzel01:_squeez_states_bose_einst_conden}.
This dynamics is governed by the time scale for which the system is
able to adiabatically follow the lowest energy state due to the
collisional repulsion among atoms, and this time scale may be very
long, making an experimental implementation very difficult
\cite{menotti01:_dynam_bose_einst}.

In contrast, we propose a 
fast method working in four  steps: {\it (i)} 
apply a normal splitting  to create a binomial distribution 
of atoms separated in space, {\it (ii)} make use of 
the collisional interaction in each spatial component to cause in 
few msecs a non-linear phase evolution of the different amplitudes,
{\it (iii)} reflect the atoms so they again overlap in space,
and {\it (iv)} remix them on a second
beam splitter, so that the resulting separate components
are populated by a sub-binomial distribution. 

In Sec. II, we present a theoretical analysis of the interferometer
applied to a Bose-Einstein condensate. In Sec. III, we discuss
the results ideally obtained with our proposal, and in Sec. IV,
we analyze the effect of a mismatch in overlap in the
interferometer.

\section{Bragg interferometer}

\begin{figure}
  \resizebox{7.5cm}{!}{\rotatebox{0}{
      \includegraphics{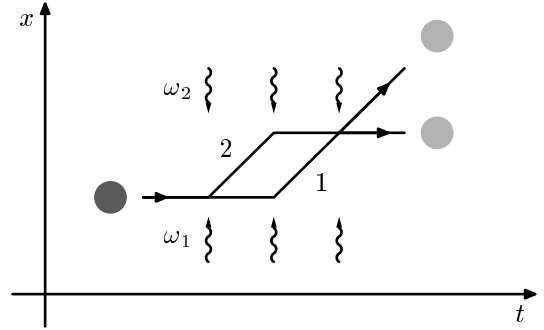}}}
  \caption{Atom-interferometer realized via Bragg scattering of the condensate.}
  \label{fig:bragg}
\end{figure}

The principle of the proposal is sketched in Fig.~\ref{fig:bragg}.,
where the vertical arrows indicate Bragg diffracting laser fields with
appropriate detunings and phases.  We suggest to use Bragg diffraction
to split and to recombine the atomic clouds because this method has
already been succesfully demonstrated in experiments.

Our proposal combines ideas for light squeezing in a Kerr-nonlinear
interferometer~\cite{kitagawa86:_number_kerr} with ideas for spin
squeezing of two-component condensates using the collisional
interactions among the atoms
\cite{soerensen01:_many_bose,poulsen01:_posit_p_bose}. In a single
condensate, the collisional interaction leads to an effective
quadratic term in the Hamiltonian proportional to $N(N-1)$, where $N$
is the total number of atoms. That term affects only the global phase
of the condensate, but if the condensate is divided in two components
with no mutual interaction, it is replaced by $n_1(n_1-1)+n_2(n_2-1)$,
having a significant effect on a superposition of states with
different populations $n_1$ and $n_2$ in the two components.  The lack
of interaction between the components could be due to a tuning of the
collisional properties of the atoms, or, as suggested in
\cite{poulsen01:_posit_p_bose}, due to a spatial separation of the
atoms.  In \cite{poulsen01:_posit_p_bose} we carried out simulations
with the so-called positive-P method which retains the full many-body
character of the problem.  Our calculations were well accounted for by
a simple two-mode description, and we shall therefore apply such a
description in the following analysis.

\subsection{Internal and external dynamics}
The atoms are initially in a single component BEC, {\it i.e.}, all
atoms populate the same one-particle wave function $\psi(\vec
r)$. Counterpropagating laser beams along the z-direction with a
frequency difference around 100 kHz are applied to the atomic cloud
and cause diffraction of the atoms. In the moving frame of the
optical standing wave pattern, Bragg diffraction conserves kinetic
energy and the atoms coherently populate two components at the
incident momentum $-\hbar k$ (corresponding to zero momentum in the
laboratory frame) and at $\hbar k$, differing by twice the photon
momentum. For suitably choosen parameters the diffraction process is
fast and interactions can be ignored. The diffraction is then a linear
process and each atom is put in a superposition of remaining at rest,
and having received twice the photon momentum. Allowed to propagate
freely the atoms will coherently populate two spatially separated
regions of space after less than a $msec$. If we neglect the role of
interactions in this short phase, we can write the state of the system
\begin{eqnarray}
|\Psi\rangle = \frac{1}{\sqrt{2^N}}\sum_{n_1=0}^N
\sqrt{\binom{N}{n_1}}
|n_1;\phi_1(\vec r)\rangle|n_2;\phi_2(\vec r)\rangle
\label{state}
\end{eqnarray}
where $|n_i;\phi_i(\vec r)\rangle$ denotes a state with $n_i$ atoms
populating the spatial wave funtion $\phi_i(\vec r)$; $n_2=N-n_1$.
In the subsequent dynamics, the wave functions $\phi_i(\vec r)$ evolve
with time, and we describe this evolution with the Gross-Pitaevskii
equations
\begin{eqnarray}
i\hbar \partial_t \phi_i = (\hat{h}_i + \frac{Ng}{2}|\phi_i|^2)\phi_i
\label{gp1}
\end{eqnarray}
where $\hat{h}_i$ is the single particle hamiltonian (the atoms may be
free, in which case $\hat{h}_i$ equals the kinetic energy operator,
they may fall under gravity, or they may be trapped in a weak trapping
potential) and $g= 4\pi\hbar^2 a_s/m$ is the collisional interaction
strength ($a_s$ is the $s$-wave scattering length). 

To compute the effect of collisions on the spatial dynamics it is an
adequate approximation to assume that half of the atoms $N/2$ populate
each component. A more precise account of the interactions will be
needed, however, when we examine the details of the evolution across
the binomial variation in $n_i$. Effectively, the functions $\phi_1$
and $\phi_2$ define two modes for the atoms with creation and
annihilation operators $a_i^\dagger$ and $a_i$, and the dynamics
associated with the distribution of atoms among the modes is accounted
for by a two-mode Hamiltonian:
\begin{eqnarray}
H=\sum_{i=1,2} g{\mathcal{I}}_i
\left(
   \frac{1}{2}(a_i^\dagger)^2a_i^2-\frac{N}{2}a_i^\dagger a_i
\right)
\label{ham2}
\end{eqnarray}
where ${\mathcal{I}}_i=\int d^3r |\phi_i|^4$, is determined from the
solution of Eq.(\ref{gp1}). It is not strictly correct to separate the
spatial and the population dynamics, especially not if the condensates
are strongly interacting, which suggests a variation of the preferred
mode function $\phi_i$ over the binominal distribution of $n_i$ around
$N/2$. In the Thomas-Fermi regime, this variation has been taken into
account within a broader ansatz for the state of a similar
two-component system suggesting, however, that this effect is taken
well care of if ${\mathcal{I}}_i$ is multiplied by a numerical factor
of order unity~\cite{sinatra00:_binar_mixtur_bose_einst} .

\subsection{Collective spin picture}
\label{sec:coll_spin_pic}
\begin{figure}[tbp]
  \resizebox{7.5cm}{!}{\rotatebox{0}{
      \includegraphics{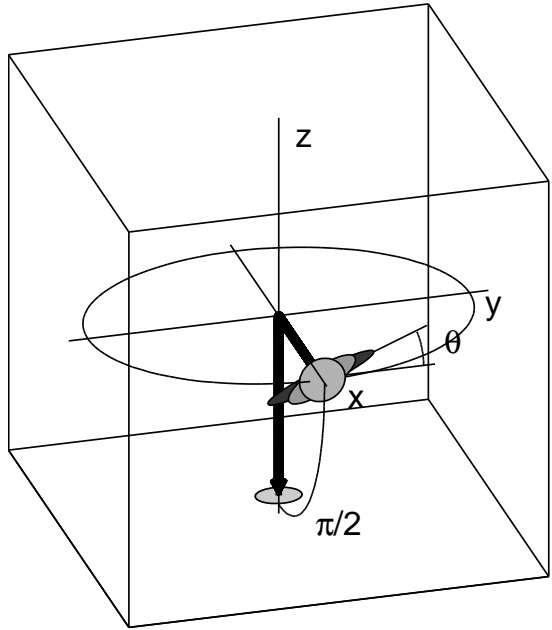}}}
  \caption{Phase- and population dynamics in the collective spin picture. 
    The $z$-component $S_z$ represents the population difference
    between the two spatial components, and the initial single
    component condensate starts out on the south pole with equal
    uncertainty in the $S_x$ and $S_y$ spin components. The
    preparation of the state with all atoms in a given superposition
    of the spatial modes 1 and 2 is represented by a simple rotation
    of the spin vector.  The first Bragg pulse is a $\pi/2$ pulse and
    the spin is rotated to the equator of the Bloch sphere. By
    convention we choose the rotation axis to be the $y$-axis so the
    mean spin ends up along the $x$-axis. The $S_z^2$ term in
    Eq.~(\ref{eq:h_spin}) now deforms the uncertainty elipse while
    leaving the mean spin along $x$. This deformation stretches the
    uncertainty elipse in the horisontal direction while
    $\mbox{Var}(S_z)$ is kept constant. To obtain reduced fluctuations
    in the populations of the two spatial modes, i.e. in $S_z$, a
    rotation by the angle $\theta$ shown in the figure is applied by a
    final Bragg pulse, chosen with approriate phases so that the
    rotation is around the direction of the mean spin.}
  \label{fig:bloch}
\end{figure}

The Hamiltonian (\ref{ham2}) will cause a quadratic (with $n_i$)
evolution of the phase of the expansion coefficients in the state
(\ref{state}). This evolution is conveniently described by
defining the collective spin operators
\begin{eqnarray}
S_+ = a_2^\dagger a_1,\ S_-= a_1^\dagger a_2, S_z = (a_2^\dagger a_2-
a_1^\dagger a_1)/2.
\label{spin}
\end{eqnarray}
The Hamiltonian (\ref{ham2}) can now be written:
\begin{eqnarray}
H=\hbar\chi S_z^2+f(N)
\label{eq:h_spin}
\end{eqnarray}
with $\hbar\chi\equiv g {\mathcal{I}}$, where we have assumed for
simplicity ${\mathcal{I}}_1={\mathcal{I}}_2\equiv{\mathcal{I}}$.  The
$f(N)$ term gives an overall phase to the system which we can neglect
while the quadratic term $S_z^2$ has been studied in detail by
Kitagawa and Ueda \cite{kitagawa93:_squeez}, who pointed out that
precisely this Hamitonian leads to squeezing, i.e., reduction of the
variance of a spin component orthogonal to the direction of the mean
spin as illustrated in Fig.~\ref{fig:bloch}.

Starting with the state (1), a spin coherent state with
the macroscopic spin pointing along the $x$-axis, 
the Hamiltonian (5) produces a state which retains a large
value of $\langle S_x\rangle$ and vanishing $\langle S_{y,z}\rangle$.
The initial variance $\mbox{Var}(S_y)=N/4$ changes due 
to the interaction, and one identifies~\cite{kitagawa93:_squeez}
a particular component $S_{\theta}=\cos\theta \;S_z - \sin\theta \;S_y$
for which the variance is minimal. In terms of $A=1-\cos^{N-1}(2\chi
t)$ and $B=4\sin(\chi t)\cos^{N-2}(\chi t)$ the angle $\theta$ is
given by
\begin{equation}
\label{eq:ueda_theta}
\theta= \frac{1}{2}\arctan \frac{B}{A}
\end{equation}
and the minimal variance is given by
\begin{equation}
  \label{eq:ueda_ds2}
  \mbox{Var}(S_{\theta}) =
  \frac{N}{4}\left\{
    1-\frac{N-1}{4}\left[
      \sqrt{A^2+B^2}-A
    \right]
  \right\}.
\end{equation}

This spin squeezing is equivalent to a collapse of the relative phase,
smeared out by the interaction
\cite{javanainen97:_phase_phase_diffus_split_bose_einst_conden,sinatra00:_binar_mixtur_bose_einst,moelmer98:_phase_bose_einst}.
It is important to notice that the phase collapse is not a decoherent
process; the system is still in a pure state, and by a simple unitary
operation described below (a spin rotation, cf. Fig.
(\ref{fig:bloch})) it can be transformed to a state with the desired
reduced population fluctuations.  We focus on population fluctuations,
but it is worhwile mentioning that it is equally possible to obtain a
state where the relative phase of the two components has become well
defined.

The Hamiltonian (5) applies whenever the two components are spatially
well separated, and we let the atomic clouds separate for a time
interval of duration $T$ (typically some $msecs$). We then apply a new
Bragg diffraction pulse to induce a complete transfer between states
with momenta $\pm \hbar k$ in the moving frame, {\it i.e.} $0$ and $2\hbar
k$ in the laboratory frame, so that the two components now approach
each other.  This operation is carried out on the two spatially
separated components, and it is accounted for by inclusion of the
interaction with the Bragg fields in $\hat{h}_i$ in Eq.(2). Note that
with the convention that the collective spin refers to the spatially
separated components, this Bragg pulse has no effect on the values of
the spin components; the solutions $\phi_i(\vec r,t)$ constitute a
'rotating frame' for our calculation of populations and coherences.

After a second time interval of duration $T$ with free evolution of
the two condensate components, they again overlap in space.  At this
point it is possible to observe interference fringes in the spatial
density
profile~\cite{andrews97:_obser_inter_between_two_bose_conden,simsarian00:_imagin_phase_evolv_bose_einst_conden_wavef}.
Note that although Fig.2 is suggestive that spin measurements and
rotations can be carried out at any time, it is only when the two
spatial components overlap, the spin variables represent physically
accessible quantities.  At the time of full overlap we can apply a
final beam splitter Bragg pulse to recombine the two components into
two new output components:
\begin{equation}
  \begin{split}
    \label{eq:bragg}
  U_{\mathrm{Bragg}} &=
  \exp \left[ i\alpha\left(
      \cos\phi\;S_x+\sin\phi\;S_y\right)\right]\\
  &= \begin{bmatrix}
    \cos\frac{\alpha}{2} & i\sin\frac{\alpha}{2} e^{-i\phi} \\
    i\sin\frac{\alpha}{2} e^{i\phi}& \cos \frac{\alpha}{2}
  \end{bmatrix} 
  \end{split}  
\end{equation}
Here we have written the Bragg evolution operator in the spin picture
and as a $2\times2$ matrix giving the mode annihilation operators
$b_1$ and $b_2$ after the pulse in terms of the operators $a_1$ and
$a_2$ before the pulse.  We are interested in creating a state with
two separate components with sub-binomial counting statistics, that
is, we want to chose $\phi$, the phase of the pulse, and $\alpha$, the
duration, such that the counting statistics of $b_1$ and $b_2$ will
benefit from the correlations in $a_1$ and $a_2$. In the spin picture
it is clear that this operation should be a rotation around the
mean spin to align the squeezed direction with the $z$-axis which
represents population differences.  If we denote by $S_{\alpha}$ the
spin component that the last Bragg pulse rotates into the
$z$-direction then from the definition~(\ref{spin}) we get the mean
and variance of the populations of one of the output beams
\begin{eqnarray}
  \label{eq:pop_dif}
  \langle 
  n_{\mathrm{det}} \rangle =\frac{1}{2}N+\langle S_\alpha 
  \rangle \nonumber \\
  \mbox{Var}(n_{\mathrm{det}}) = \mbox{Var}(S_\alpha).
\end{eqnarray}
If $\phi=0$ we rotate around the $x$-axis and the angle $\alpha$
should ideally be choosen equal to $\theta$ of Fig.~\ref{fig:bloch}
and Eq.~(\ref{eq:ueda_theta}). With no spin squeezing one finds
$\mbox{Var}(n_{\mathrm{det}})=N/4$ in agreement with the initial
binomial distribution.

From Eq.~(\ref{eq:bragg}) it is clear that the distinction between
different rotation axes, all lying in the $xy$-plane is equivalent to
the one between complex phases $\phi$ of the coupling amplitude, {\it
  i.e.}  the relative phase of the two counterpropagating laser beams
creating the moving standing wave. Alternatively, we see that the
location of the Bragg diffraction pattern, due to the translational
properties of momentum eigenstates, controls precisely the phase of
the coupling.  In Ref.~\cite{poulsen01:_posit_p_bose}, we learned that
the mean spin and the rotation angle are predicted with adequate
precision in the simple model~(\ref{eq:h_spin},\ref{eq:ueda_theta}),
so that application of Bragg pulses with these variables should
suffice to yield substantial reduction of the population fluctuations
of the two atomic outputs. It is also straightforward to optimize the
parameters in experiments. The phases of the diffracting lasers are
adjusted so that the two output beams have the same mean occupancy,
independently of the duration of the coupling. The solid curve in
Fig.~\ref{fig:effect_of_c} shows the squeezing factor
$(N/4)/\mbox{Var}(n_{\mathrm{det}})$ (large for strong squeezing) as a
function of $\alpha$. With a spin squeezed sample the variance of the
populations show a strong dependence on the duration of the last Bragg
pulse, which should therefore be adjusted to identify the output with
minimal fluctuations.

\section{Expected results}
\label{sec:expected_results}
Let us turn to the discussion of the experimental feasibility of our
proposal. Splitting and recombination of one-component condensates
have been done in several laboratories, and the coherence properties
have been verified and explicitly utilized in a number of imaging
experiments: our proposal follows closely the experiments at NIST,
where the phase variation of the condensate gives rise to a density
variation in the recombined output
\cite{denschlag00:_gener_solit_phase_engin_bose_einst_conden}. These
experiments have been done in the limit of interacting condensates
which is of course essential for our proposal:
In~\cite{simsarian00:_imagin_phase_evolv_bose_einst_conden_wavef} the
mean field repulsion was observed and it was shown not to prevent a
nearly perfect overlap of the spatial modes at the recombination, and
in~\cite{denschlag00:_gener_solit_phase_engin_bose_einst_conden} a
soliton was imprinted in one of the components to be subsequently
detected in the output. We note, that also the phase variation around
quantized vortices has been studied by similar interference imaging,
both in the case where a vortex was prepared prior to the splitting
and a dislocation appears in the interference fringes
\cite{chevy01:_inter_bose_einst}, and in the case where a condensate
was split and a vortex was subsequently created by stirring only one
component \cite{inouye01:_obser_bose_einst}.

To assess the strength of the squeezing interaction we need to
evaluate the parameter $\chi$ of Eq.(\ref{eq:h_spin}). In the
Thomas-Fermi approximation, valid for an interaction dominated
condensate, one finds the simple result:
\begin{equation}
\label{eq:phi4}
{\mathcal{I}}=\int |\phi|^4 d^3r=\frac{10}{7}\frac{1}{V_{\mathrm{TF}}}
\end{equation}
where $V_{\mathrm{TF}}\equiv (4\pi/3) R_x R_y R_z$ with $R_i$ the
Thomas-Fermi radius of the condensate in the $i$-direction.  With the
parameters of the experiments with condensates, being split and
recombined by Bragg diffraction at NIST,
\cite{denschlag00:_gener_solit_phase_engin_bose_einst_conden}, one
finds $\chi=4.7\times 10^{-4} s^{-1}$ or a factor of $\sim 4$
reduction in the population variances in the 2 $msec$ duration of the
experiment.  This result is improved to a factor of more than 12 by
increasing the duration to 4 $msec$. The theoretical maximal squeezing
factor $\sim (1/3)(N/6)^{-2/3}\sim 1.3 \times
10^5$~\cite{kitagawa93:_squeez} would require the components to be
separated for $\sim 6^{1/6}\chi N^{-2/3} \sim 200 msec$ assuming a
constant $\chi$ throughout the experiment. In reality the condensate
components expand if the trap is turned off during the experiment and
the continuous reduction of $\chi$ over longer times according to
Eq.(\ref{eq:phi4}) would have to be determined by an integration of
the Gross-Pitaevskii equation (2).

\section{Effect of imperfect overlap}
As a final issue we wish to discuss the matching of the spatial wave
functions of the two components. After the physical interactions
described above, one will detect a number of atoms with momenta around
zero, and a number of atoms with momenta around $2\hbar k$, and the
analysis shows that these numbers will fluctuate less than if they
were given by a binomial distribution. This reduction is due to the
correlations between the atoms following the lower and the upper paths
in the interferometer, and it is hence important that atoms from the
lower path, being diffracted by the last Bragg pulse, occupy the same
spatial state as the undiffracted component of the upper path, cf.
Fig.~\ref{fig:bragg}. In the extreme case where these two
contributions are orthogonal and distinguishable, there will be no
squeezing effect at all, and in the case where they overlap 100 \%,
the squeezing is given by the above simple analysis.

Atoms leaving the interferometer at zero momentum can in general be
described by a superposition of $\phi_2$, the undiffracted upper path
wave function, and a state $\phi_{2\!\perp}$ orthogonal to $\phi_2$.
This gives rise to two orthogonal modes which both contribute to the
detected number of atoms around zero momentum. We can express the
field annihilation and creation operators $b_2,\ b_{2\!\perp}$ and
$b_2^{\dagger},b_{2\!\perp}^{\dagger}$ for these two modes in terms of
the operators $a_{1},a_{2}$ and $a_{1}^{\dagger},a_2^\dagger$ given
above for the modes prior to the last Bragg pulse {\it and} operators
for two additional 'vacuum' modes $a_{1\!\perp},a_{2\!\perp}$ and
$a_{1\!\perp}^\dagger,a_{2\!\perp}^\dagger$, which are necessary to
ensure unitary of the beam splitter pulse. Letting $c$ denote the
overlap of the (normalized) diffracted lower path component with the
upper path undiffracted component and $s$ the overlap with the
orthogonal complement 
\begin{equation}
  \label{eq:def_c}
  e^{-2ikx}\phi_1(\vec{r}) = c \phi_2(\vec{r})+s \phi_{2\!\perp}(\vec{r})
\end{equation}
($|s|^2+|c|^2=1$). In terms of $\alpha$ and $\phi$ of the last Bragg
pulse~(\ref{eq:bragg}) we have:
\begin{align}
  \label{eq:b}
  b_2 =&\cos\frac{\alpha}{2} \; a_2
  + i  \sin\frac{\alpha}{2} e^{-i\phi}
  \left(
    c  \; a_1 
    + s \; a_{1\!\perp}
  \right)
  \\
  b_{2\!\perp} =& \cos\frac{\alpha}{2} \; a_{2\!\perp}
   +i \sin\frac{\alpha}{2} e^{-i\phi}
  \left(
    s^* \; a_1 
    - c^* \; a_{1\!\perp}
  \right)
  .
\end{align}
Focusing on $b_2$ we see that for $|c|<1$ we can
still control the mixing and pick the squeezed combination of $a_1$
and $a_2$.  However, some contributions from the unoccupied modes have
been introduced, and the situation is similar to the detection of
squeezed light: A non-unit overlap $\nu$ between the squeezed mode of
light and a detector mode as, e.g., modelled by a beam splitter
transmitting a fraction $|\nu|^2$ of the squeezed light to the
detector, leads to admixture with a vacuum field and the deterioration
of the squeezing,
$\mbox{Var}(X)_{\mathrm{det}}=|\nu|^2\mbox{Var}(X)_{sq}+(1-|\nu|^2)$,
where the $X$ is the scaled quadrature field component with unit
variance in the vacuum state.

In a real experiment it is difficult to count the atoms in mode
$\phi_2$ exclusively. It is more realistic that one will only
have access to the total number of atoms in $\phi_2$ and
$\phi_{2\!\perp}$. We then have to include $b_{2\!\perp}$ in the
analysis, and if the phase of the Bragg diffraction is chosen to
compensate the phase of the complex overlap $c$, we find:
\begin{eqnarray}
  \label{eq:Ndet}
  \langle n_{\mathrm{det}} \rangle &=& 
  \langle 
  b_{2}^{\dagger}b_2+b_{2\!\perp}^{\dagger}b_{2\!\perp}
  \rangle \nonumber\\
  &=&
  \langle
  \cos\alpha \; S_{z} -|c| \sin\alpha \; S_{y} 
  \rangle 
  + \frac{N}{2}  \nonumber\\
  &=& \lambda \langle S_{\alpha'} \rangle + \frac{N}{2}  .
\end{eqnarray}
in terms of the spin component $S_{\alpha'}$ \emph{prior} to the 
last Bragg pulse. The angle $\alpha'$ is related to $\alpha$, the
rotation angle given by the duration of the last Bragg pulse, 
by $\tan\alpha'=|c|\tan\alpha$, and we have introduced the factor
$\lambda = \sqrt{|c|^2/(\sin^2\alpha'+|c|^2\cos^2\alpha')}$.  

The spin component $S_{\alpha'} \equiv \cos\alpha' S_z -
\sin\alpha' S_y$ is in a direction perpendicular to the mean spin. We
therefore have $\langle S_{\alpha'} \rangle =0$ and $\langle n_{\mathrm{det}}
\rangle = N/2$ independently of $\alpha$. The variance
of $n_{\mathrm{det}}$ is given by
\begin{eqnarray}
  \label{eq:varNdet}
  \mbox{Var}(n_{\mathrm{det}})&=&\langle n_{\mathrm{det}}^2 \rangle 
  - \langle n_{\mathrm{det}} \rangle^2 \nonumber\\
  &=& \lambda^2 \mbox{Var}(S_{\alpha'})+(1-\lambda^2) \frac{N}{4} .
\end{eqnarray}
If the overlap is complete $|c|=1=\lambda$, we can pick out the maximally
squeezed spin component to yield the minimal variance of $n_{\mathrm{det}}$.
In case of a non-perfect overlap, we see that Eq.~(\ref{eq:varNdet})
introduces a non-squeezed contribution, scaling as $N/4$, and it makes
the number fluctuations depend on the noise of the spin component
$S_{\alpha'}$ rather than of $S_{\alpha}$. We therefore want $\alpha'$
to be close to the actual direction of squeezing $\theta$ in Fig.2 and
Eq.(\ref{eq:ueda_theta}), which is obtained by choosing a Bragg pulse
with longer duration than what is optimal in the ideal
case. 

\begin{figure}[tbp] 
    \resizebox{7.5cm}{!}{\rotatebox{0}{
        \includegraphics{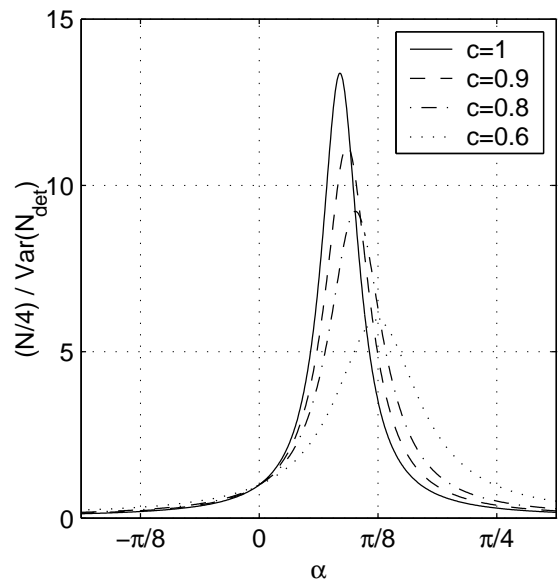}}}
    \caption{The reduction of 
      $\mbox{Var}(n_{\mathrm{det}})$ is illustrated by plotting the
      reciprocal ratio $(N/4)/\mbox{Var}(n_{\mathrm{det}})$ as a
      function of the applied Bragg pulse rotation angle for
      parameters as in Sec.~\ref{sec:expected_results} with 4 $msec$
      separation. The results are shown for different values of the
      overlap, $c=1$ (solid line), $c=0.9$ (dashed line), $c=0.8$
      (dash-dotted curve), and $c=0.6$ (dotted curve).  The value of
      $\alpha$ at the maximum of the $c=1$ curve displays the
      orientation of the squeezing elipse, i.e. $\theta$ of
      Fig.~\ref{fig:bloch}. As is apparent from the other curves, one
      may compensate for a non-perfect overlap by applying a larger
      Bragg pulse rotation.}
  \label{fig:effect_of_c}
\end{figure}

In Fig.~\ref{fig:effect_of_c} 
we show the variation of $(N/4)/\mbox{Var}(n_{\mathrm{det}})$ with
$\alpha$ for different values of the overlap $c$. When $|c|$ is reduced
the squeezing factor drops, but the curves illustrate that the results
improves if one choses a larger rotation angle $\alpha$.

Note that the mode function mismatch does not have the same
detrimental effect as in the detection of a single mode squeezed
field: the admixture of the vacuum modes, and hence the vacuum
contribution $N/4$ to $\mbox{Var}(n_{\mathrm{det}})$, does not scale
with $|s|^2=1-|c|^2$ but rather with $|s|^2\sin^2\alpha$, and for
small rotation angles this is a small number. For strong squeezing
the uncertainty elipse is almost horisontal and only rotation by a
small $\alpha$ is needed to observe squeezing in the number
fluctuations. In this case even a significantly reduced overlap does
not prevent a significant noise reduction for $n_{\mathrm{det}}$. This is
illustrated when we write the maximal squeezing factor attainable for
given $c$ and given $\mbox{Var}(S_\theta)$:
\begin{eqnarray}
  \label{eq:M}
  \mbox{Var}(n_{\mathrm{det}}) &=& \mbox{Var}(S_\theta) \left(
    1+(1-|c|^2) \frac{\frac{N}{4}-\mbox{Var}(S_\theta)} {|c|^2
      \frac{N}{4}+\mbox{Var}(S_\theta)}
  \right)\nonumber\\
  &\sim&\quad
  \frac{\mbox{Var}(S_\theta)}{|c|^2}\qquad(\mbox{Var}(S_\theta)\ll
  \frac{N}{4}).
\end{eqnarray}
(In Eq.~(\ref{eq:M}) we assume that the spin state described by
Eqs.~(\ref{eq:ueda_theta},\ref{eq:ueda_ds2}) is a minimum uncertainty
state). 

\section{Conclusion}

We have proposed to use the collisional interactions between atoms in
the separate arms of a spatial interferometer to suitably manipulate
the amplitudes on the different number state components. We analyzed
the achievements of current condensate interferometry experiments, and
we found that significant number squeezing should be found in the
interferometer output, i.e., the interferometer acts as a quantum beam
splitter for atoms with unique noise properties. Good overlap of the
output mode functions is necessary to observe interference effects in
the interferometer and it is also important for the noise reduction in
our beam splitter. We showed, however, that the overlap is not a
critical parameter and sizable noise reduction is possible with
interferometeric overlaps easily achievable in experiments.


\end{document}